\documentclass[review]{elsarticle}
\biboptions{sort&compress,numbers}
\usepackage[margin=1in]{geometry}
\usepackage{float}
\usepackage[colorlinks=false]{hyperref}\hypersetup{pdfborder={0 0 0}}
\usepackage{miller}
\usepackage{bm,amsmath,amssymb,amsthm}
\usepackage{enumitem}
\usepackage{mathalfa}
\usepackage{cancel}
\usepackage{subfigure}\usepackage{graphicx}
\usepackage[margin=1in]{geometry}
\usepackage{booktabs,dcolumn,caption}
\newcolumntype{d}[1]{D{.}{.}{#1}} 
\renewcommand{\ast}{{}^{\textstyle *}} 

\newcommand\U{\bm{p}}

\newtheorem{definition}{Definition}

\begin{document}

\begin{frontmatter}
\title{Optimal transportation of grain boundaries: A forward model for predicting migration mechanisms}
\author[cmu]{Ian Chesser}
\ead{ichesser@andrew.cmu.edu}
\cortext[cor1]{Corresponding author}
\author[cmu]{Elizabeth Holm}
\ead{eaholm@andrew.cmu.edu}
\author[uccs]{Brandon Runnels\corref{cor1}}
\ead{brunnels@uccs.edu}

\address[cmu]{Department of Materials Science and Engineering, Carnegie Mellon University, Pittsburgh, PA 15213}
\address[uccs]{Department of Mechanical and Aerospace Engineering, University of Colorado, Colorado Springs, CO, United States}

\begin{abstract}

It has been hypothesized that the most likely atomic rearrangement mechanism during grain boundary (GB) migration is the one that minimizes the lengths of atomic displacements in the dichromatic pattern. In this work, we recast the problem of atomic displacement minimization during GB migration as an optimal transport (OT) problem. Under the assumption of a small potential energy barrier for atomic rearrangement, the principle of stationary action applied to GB migration is reduced to the determination of the Wasserstein metric for two point sets. In order to test the minimum distance hypothesis, optimal displacement patterns predicted on the basis of a regularized OT based forward model are compared to molecular dynamics (MD) GB migration data for a variety of GB types and temperatures. Limits of applicability of the minimum distance hypothesis and interesting consequences of the OT formulation are discussed in the context of MD data analysis for twist GBs, general $\Sigma 3$ twin boundaries and a tilt GB that exhibits shear coupling. The forward model may be used to predict atomic displacement patterns for arbitrary disconnection modes and a variety of metastable states, facilitating the analysis of multimodal GB migration data.

\end{abstract}

\end{frontmatter}



\section{Introduction}

Grain boundaries (GBs) are mobile planar defects in polycrystalline materials that comprise a large structural and dynamic phase space. Controlling GB structure and mobility is an important challenge during the processing of metallic and ceramic materials \cite{petch1953cleavage,chiba1994relation,randle2010grain,sangid2013physics,chookajorn2012design,shimada2002optimization,rollettrecrystallization,zhang2013stress,rupert2009experimental,li2013incoherent,jin2014annealing,lqcke1992texture}. Since Transmission Electron Microscopy (TEM) experiments cannot yet track the motion of individual atoms during GB migration \cite{merkle2004situ,wei2021direct}, Molecular Dynamics (MD) simulations offer the most detailed accounts of GB migration mechanisms currently available, albeit at small length and timescales rarely exceeding 100 nm and 10 ns. MD studies of GB migration have revealed wide variation in the thermal and mechanical behavior of GBs with crystallography, temperature and constraints \cite{olmsted2009survey,han2018grain,thomas2017reconciling,thomas2019disconnection,chen2020temperature}. Mobility in FCC Ni bicrystals was found to vary over four orders of magnitude with GB type, with some GBs moving near the speed of sound in Ni and others remaining almost stationary \cite{olmsted2009survey,o2016exploration}.  Ongoing challenges in GB migration science include 1) the identification of unit mechanisms and structural degrees of freedom that correlate to specific migration behavior (e.g. high mobility) and 2) the development of continuum models that faithfully coarse grain defect mechanics learned at the atomic scale.

A question central to modeling GB migration is whether there exists a unique correspondence between GB structure and GB motion. The answer to this question depends on the specified resolution of GB structure as well as thermodynamic variables (temperature, driving force, etc). It is generally accepted that there is no unique correspondence between macroscopic GB geometry (misorientation and boundary inclination) and GB motion. Every macroscopically defined GB has a countably infinite set of allowed interfacial defects with both step and dislocation character called disconnections. Often, distinct disconnection modes can be distinguished by a scalar geometric parameter called the coupling factor that describes the ratio of sliding and migration. Energy barriers for disconnection nucleation and motion are key input parameters to continuum models for coupling factor and grain boundary mobility \cite{chesserout,han2018grain,chen2020temperature}. Multiple disconnections modes with closely spaced energy barriers may compete at finite temperature, leading to interesting phenomena such as GB roughening (the onset of significant motion at a finite temperature) and zig-zag mode switching under constraints \cite{thomas2017reconciling}. These phenomena violate the assumption of unique correspondence between macroscopic geometry and migration mechanism. 

If one refines the definition of GB structure to include the atomic structure of the interface, more possible GB motion mechanisms emerge. Metastable GB domains known as complexions are associated with different microscopic degrees of freedom for fixed macroscopic parameters. For pure materials, microscopic degrees of freedom include a rigid translation between grains and GB density \cite{Han2016, frolov2013structural,hickman2017extra}. GB motion is sensitive to microscopic degrees of freedom, as evidenced by the dependence of mobility and shear coupling factor on complexion type \cite{krause2019review,frolov2014effect}. Moreover, GB diffusion and rearrangement events associated with changes in metastable GB structure are known to proceed through a variety of kinetic pathways \cite{suzuki2005atomic,frolov2013structural,sharp2018machine}. There is a need to develop computationally efficient models which optimize kinetic quantities such as mobility over both macroscopic and microscopic degrees of freedom. 

It has been proposed that, given a disconnection nucleation event at fixed microscopic degrees of freedom, a unique correspondence can be made between GB structure and motion \cite{hirth2016disconnections}. In a review of disconnection related phenomena including martensitic transformations, twin nucleation and GB migration, Hirth set out following qualitative criteria to determine collective atomic displacements during disconnection nucleation and motion: 1) minimum possible Burgers vector to minimize elastic energy and the Peierls barrier and 2) minimum shuffle lengths to minimize activation energy and possible entropic contributions \cite{hirth2016disconnections}. Remarkably, these somewhat simple geometric considerations have led to insights into twin morphology in complex minerals for which reliable interatomic potentials are currently unavailable \cite{hirth2019topological,hirth2020disclinations}. Extending Hirth's second condition, it has been hypothesized that the selected displacement pattern during GB migration corresponds to that which minimizes the total distance traveled by the atoms.
We refer to this here and subsequently as the \textit{min-shuffle hypothesis}. Displacement minimization has provided useful in related contexts, including prediction of orientation relationships (habit planes) during martensitic transformations \cite{therrien2020minimization} and identification of fast transition pathways for diffusionless phase transformations \cite{stevanovic2018predicting}. The validity of the min-shuffle hypothesis has not been thoroughly tested for a wide range of GBs. In light of this, in this work we develop methods to test the min-shuffle hypothesis against molecular dynamics (MD) data for GB structures with varying macroscopic/microscopic structure at a variety of temperatures. 

A contribution of this work is reframing the min-shuffle hypothesis as an optimal transport (OT) problem. Within the framework of OT and stationary action, we put quantitative bounds on when the min-shuffle hypothesis is expected to hold. We develop a forward model for predicting probability distributions over allowed transformations in the dichromatic pattern that captures variation in macroscopic and microscopic GB geometry. Regularization in the forward model enables generation of competing transformation pathways and serves as a proxy for temperature in MD simulations, where multiple mechanisms may compete simultaneously. A series of targeted MD migration simulations are chosen in this work to illustrate the limits of applicability of the min-shuffle hypothesis and OT model.  

As a mathematical framework for interpolating between arbitrary probability distributions, optimal transportation has seen much activity in the last decade. 
In 1999, Benamou and Brenier  discovered a surprising fluid mechanics reformulation of the OT problem that connects the underlying mathematical principle to pressureless potential flow \cite{benamou2002monge}. 
This was later used by Li, Habbal and Ortiz to develop an optimal transportation meshfree method for solid and fluid flows \cite{li2010optimal}.
Around the same time, Jordan {\it et al.} showed that the Fokker-Planck equation (of relevance in molecular dynamics) is the gradient flow of a thermodynamic free energy with respect to the Wasserstein distance \cite{jordan1998variational}. 
More recently, it has been proven that the Wasserstein distance provides a lower bound for work dissipation for certain Langevin stochastic processes in finite time \cite{chen2020stochastic}. 
OT has been used in materials science to develop the lattice-matching model for grain boundary energy \cite{runnels2016analytical}. 
Grain boundary migration is particularly conducive to treatment with optimal transport, as both the initial and final states are known.
Indeed, the process of grain boundary motion itself may be regarded as a transport mechanism that is optimal with respect to some transportation cost functional.
In this work, we take an approach inspired by \cite{benamou2002monge} to connect the kinetic energy of collective atomic displacements during grain boundary migration to the Wasserstein distance between point sets.  
We begin with classical mechanics of atomic motion, and show that the solution to the classical mechanics equations are reducible to the computation of a Wasserstein metric, given sufficiently fast atomic motion.

The paper is structured in the following way. 
In Section~\ref{sec:theory} we provide a physically-motivated justification for the min-shuffle hypothesis in the context and language of optimal transport.
The optimal transport formulation is modified using the Kantorovich and entropic regularizations to improve solvability and account (heuristically) for competing shuffle patterns at finite temperature. In Section~\ref{sec:methodology} we discuss the implementation of the computational algorithm and the molecular dynamics results that are used for comparison. In Section~\ref{sec:results_and_discussion} we demonstrate how the forward model can be used to test the min-shuffle hypothesis against MD data for several GBs. We show that non-optimal displacements play an important role in high temperature, high driving force or high $\Sigma$ migration of twist GBs, assessing the applicability of the min-shuffle hypothesis and OT model predictions in this context. Next, we consider high mobility $\Sigma 3$ GB migration and argue that this case is close to satisfying the OT model assumptions. We discuss the important role of the microscopic translation vector in determining the boundary plane anisotropy of shuffling patterns for $\Sigma 3$ GBs. Finally, we demonstrate via analysis of shear coupled migration of a symmetric tilt GB that optimal shuffles may be predicted for arbitrary disconnection modes that closely match MD data. Displacement pattern predictions for multiple disconnection modes are used to interpret high temperature, zero shear migration of the $\Sigma 5 \hkl{310}$ GB.

\section{Theory}
\label{sec:theory}

In this section we develop a forward model for predicting grain boundary shuffle distributions, as well as the optimal microscopic translation vector.
The model is based in the Lagrangian formulation of classical mechanics.
It is subsequently shown that the principle of stationary action (in the context of boundary migration) is equivalent to the determination of the Wasserstein metric for two point sets, provided that the potential energy contribution is sufficiently small.
This connects the ideas of Newtonian mechanics and boundary migration to the well-studied field of optimal transport theory.
The result is a computationally efficient, predictive model for grain boundary atomic shuffles and microscopic translation vectors.

We begin by establishing formal definitions for atomic positions before and after grain boundary motion.
We adopt the convention that the boundary moves so as to transform lattice 1 into lattice 2, and define the reference atomic sites corresponding to the ideal crystallographic configration before and after motion:
\begin{definition}[Reference atomic sites]
    $\mathcal{X}\subset\mathbb{R}^3$ and $\mathcal{Y}\subset\mathbb{R}^3$ are the reference atomic sites for lattice 1 and 2 respectively, if:
    \begin{enumerate}[nosep]
        \item $\mathcal{X}$ and $\mathcal{Y}$ posess the same point group symmetries of the lattice
        \item There exists an isometry $R$ such that $\mathcal{Y}=R\mathcal{X}$.
        \item $\mathcal{X}\cap\mathcal{Y}$ is the coincident-site lattice.
        \item $\mathcal{X}\cup\mathcal{Y}$ is the dichromatic pattern.
    \end{enumerate}
\end{definition}

We note that only boundaries with a rational misorientation angle are considered, which is equivalent to requiring that $\mathcal{X}\cap\mathcal{Y} > 1$ for cubic lattices. 
\begin{definition}[Unit cells]
    $B,b\subset\mathbb{R}^3$ are unit cells for lattices 1 and t, respectively, if the following is true:
    \begin{enumerate}
        \item Periodicity: the quotient groups $\bm{X}=\mathcal{X}/B,\bm{Y}=\mathcal{Y}/b$ are partitions of $\mathcal{X},\mathcal{Y}$.
        \item Compatibility: There exists $\bm{F}_{GB}\in SL_3(\mathbb{R})$  such that $[\bm{F}_{GB}-\bm{I}]\bm{N}=\bm{0}$ for boundary plane normal $\bm{N}$, and $b = \bm{F}_{GB}B$.
        \item No-boundary: $\partial B= \partial b = 0$.
        Opposite edges are glued together so that an atom moving out through one face re-appears on the other face.
    \end{enumerate}
    Note: in n dimensions, $B,b$ are topologically equivalent to each other and to the n-torus $(S^1)^n$.
\end{definition}

\begin{figure}[t]
  \centering
  \begin{minipage}{0.5\linewidth}
    \includegraphics[width=\linewidth]{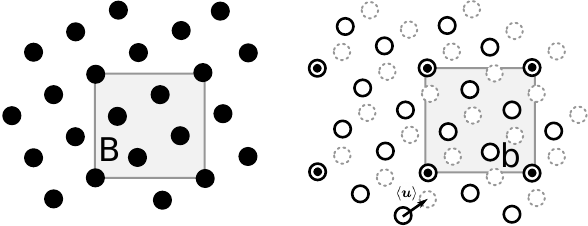}
  \end{minipage}
  \caption{
    Definitions: 
    \{\protect\includegraphics{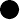}\}$ = {\mathcal{X}}$, 
    \{\protect\includegraphics{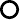}\}$ = {\mathcal{Y}}$,
    \{\protect\includegraphics{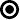}\}$ = {\mathcal{X}}\cap{\mathcal{Y}}$=CSL,
    \{\protect\includegraphics{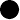}\}$ = {\mathcal{Y}}+\U$.
  }
  \label{fig:definitions}
\end{figure}

Figure~\ref{fig:definitions} illustrates the atomic sites and unit cells as defined above.
The above two conditions require that $B,b$ be prismatic with at least one shared plane corresponding to the boundary plane.
For every boundary with finite $\Sigma$, it is possible to select a $B$ and a $\bm{F}_{GB}=\bm{I}$ so that $B=b$.
This corresponds to the special case of no-shear transformation in which there is no enforced shear coupling.
Note, however, that this does not preclude shear coupling. 

In general, boundary motion carries a microscopic translation, meaning that $\mathcal{X}\cancel{\mapsto}\mathcal{Y}$ in general.
If the transformation carries an average translation $\U$, then $\mathcal{X}\mapsto\mathcal{Y}+\U$.
It will be shown that for any bicrystal, there is value for $\U$ that is optimal for facilitating migration, which we will denote $\U^*_{M}$.
This shift vector is not necessarily the same as that corresponding to the grain boundary energy-minimizing shift, which we will call $\U^*_{GB}$.
Predicting the optimal translation vector $\U$, and in particular the distinction between $\U^*_M$ and $\U^*_{GB}$, is of interest in this work.

The finite quotient groups $\bm{X},\bm{Y}$ correspond to unique atomic positions within the periodic unit cell, and we further define the group $\bm{x} = (\mathcal{Y}+\U)/b$.
The coset representatives for the quotient groups are
\begin{align}
  X &= \{X_i=\bm{X}_i\cap B, \bm{X}_i\in\bm{X} \}
  &Y &= \{Y_i=\bm{Y}_i\cap b, \bm{Y}_i\in\bm{Y} \}
  &x &= \{x_i=\bm{x}_i\cap b, \bm{x}_i\in\bm{x} \},
\end{align}
which represent sets of atoms that are inside $B$ (for $X$) and $b$ (for $Y,x$).
It is straightforward to show that $|X|=|Y|=|x|$, therefore, there exist bijective mappings $X\mapsto Y,X\mapsto x$ corresponding to a permuation $\sigma$.

Finally, let $\Gamma_\sigma(X,x)$ be the collection of the sets of continuous trajectories from $\bm{X}\to\bm{x}$ with permutation $\sigma\in\Sigma$ on $(\bm{X},\bm{Y})$ over an interval $[0,1]$.
Formally,
\begin{align}
  \Gamma_\sigma(X,x) = \{\gamma\in\{\gamma_i(t) \in C_2([0,1],\mathbb{R}^3) : \gamma_i(0) = X_i, \gamma_i(1)=x_{\sigma(i)}\}_{i=1}^n\}.
\end{align}

\subsection{Stationary action}

We are now in a position to derive a simplified, yet physically-motivated method for determining permutations (shuffling transformations) and the corresponding trajectories $\Gamma$.
We begin by considering representative atoms only, ($X,x$), with no shear coupling ($\bm{F}_{GB}=\bm{I}$); we will then generalize to $\bm{X},\bm{Y},\bm{x}$ and shear-coupled boundaries.

We assume that the atoms move classically under the influence of some interatomic potential $U(x_1,\ldots,x_n)$ and that the transformation time is bounded by some constant $\tau$.
The Lagrangian formulation of classical mechanics holds that the trajectories $\gamma^*$ are stationary points of the action functional $J:\Gamma_\sigma\to\mathbb{R}$.
which is defined generally as
\begin{align}\label{eq:princ_least_action}
  J[\gamma] = \tau\int_0^1 \mathcal{L}(\gamma_1,\ldots,\gamma_n,\dot{\gamma}_1,\ldots,\dot{\gamma}_n)\,dt,
\end{align}
where the dot indicates the time derivative.
If atoms all possess mass $m$ then the Lagrangian $\mathcal{L}$ is
\begin{align}\label{eq:lagrangian}
  \mathcal{L} = \sum_i\frac{1}{2}m|\dot\gamma_i|^2 - U(\gamma_1,\ldots,\gamma_n).
\end{align}
Setting the variational derivative of (\ref{eq:princ_least_action}) to zero recovers the classic time evolution equations for $\gamma^*$ corresponding to Newtonian physics.
The limitation of (\ref{eq:princ_least_action}) is that it is not necessarily convex, and stationary trajectories can be located at saddle points of $J$ rather than minima.

\begin{figure}
  \centering
  \includegraphics{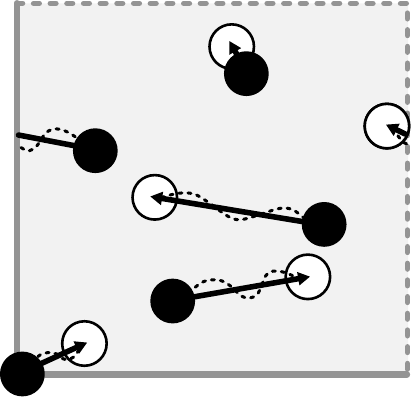}
  \caption{
    Schematic representation of transformation paths $\gamma$ decomposed into geodesics $\hat\gamma$ (solid arrows) and fluctuations $\eta$ (dashed lines).
  }
  \label{fig:gbsets_subset}
\end{figure}

We now aim to reduce the least action principle to a reduced-order model for predicting atomic shuffles.
For a given permuation $\sigma$, the distance-minimizing paths for this permuation are
\begin{align}
    \hat{\gamma}_\sigma = \underset{\gamma\in\Gamma_\sigma}{\operatorname{arg}\inf}\sum_{\gamma_i\in\gamma}\int_{\gamma_i}ds.
\end{align}
We refer to the family of trajectories  $\hat{\gamma}_\sigma$ as the geodesics on $B$ corresponding to $X,x,\sigma$.
We adopt the {\it ansatz} that the actual trajectories deviate from $\hat{\gamma}_\sigma$ by admissible fluctuations $\eta_\sigma$, where
\begin{align}
    \eta_\sigma = \Gamma_\sigma(0,0),
\end{align}
as shown in Figure~\ref{fig:gbsets_subset}.
For a given permutation and a small fluctuation from the geodesic, $J$ can thus be Taylor expanded about the geodesic $\hat{\gamma}_\sigma$ to the first order using the variational derivative:
\begin{align}
  J[\gamma_\sigma] = J[\hat{\gamma_\sigma}] + \int_{[0,\tau]}\Big(\sum_i\Big(\frac{\delta \mathcal{L}}{\delta \gamma}_i\Big)_{\gamma=\hat\gamma} \cdot \eta_i \Big)\, dt + \text{h.o.t.}
\end{align}
Substituting (\ref{eq:lagrangian}), 
\begin{align}\label{eq:recombine}
  J[\hat\gamma + \eta]
  &=
    \frac{m}{2\tau}\sum_i\,|x_{\sigma(i)}-X_i|^2 - \tau\int_{[0,1]}\Big[U(\hat{\gamma}_i,\ldots,\hat{\gamma}_n) + \sum_i\frac{\partial U}{\partial\gamma_i}(\hat{\gamma}_1,\ldots,\hat{\gamma}_n) \cdot \eta_i + \text{h.o.t.}\Big]\,dt 
\end{align}
Remarkably, the reduction of the kinetic energy term to the sum of the squared distances is exact and does not change with the addition of higher order terms.
This equivalence is the discrete analog to the optimal transport formulation of fluid mechanics by Benamou and Brenier \cite{benamou2000computational}.
Assuming small perturbations and applying nondimensionalization allows us to reduce the least action principle to the following form
\begin{align}
    \min_{\sigma\in\Sigma}\inf_{\eta\text{ admis.}}\Big[ \frac{1}{2}\sum_i\,\big|\frac{x_{\sigma(i)}-X_i}{\alpha}\big|^2 - \mathcal{D}\int_{[0,1]}\frac{U(\hat{\gamma}_i,\ldots,\hat{\gamma}_n)}
    {\langle U \rangle} \,dt \Big],
\end{align}
where $\langle U \rangle$ is an expected (possibly temperature-dependent) value for the potential term, and 
\begin{align}
    \mathcal{D} = \frac{\tau^2\langle U\rangle}{m \alpha^2}
\end{align}
is a dimensionless constant that quantifies the relative value of kinetic and potential energy contributions. 
When $\mathcal{D}$ is sufficiently small, the potential term is negligible and can be neglected.
The result is
\begin{align}
    \min_{\sigma\in\Sigma} \frac{1}{2}\sum_i\,\big|\frac{x_{\sigma(i)}-X_i}{\alpha}\big|^2.\label{eq:wasserstein}
\end{align}
In summary, in the short-time heavy-atom limit, the principle of stationary action is reducible to the Wasserstein 2-metric between lattice points. 
The statement of (\ref{eq:wasserstein}) is equivalent to that of the min-shuffle or least-distance hypotheses, but we have shown that it can be linked to the physical principle of stationary action in the low-$\mathcal{D}$ limit.

We now seek to devise an efficient algorithm for solving (\ref{eq:wasserstein}).
The calculation of (\ref{eq:wasserstein}) is the well-known Monge problem which has a computable solution with $\mathcal{O}(n!)$ complexity \cite{maretic2020wasserstein}.
This problem can be recast as a linear program by applying the Kantorovich regularization.
We replace the coset representatives $X,Y,x$ with the quotient groups $\bm{X},\bm{Y},\bm{x}$, and the permutation $\sigma$ with a bistochastic probability matrix $P\in\Pi(\bm{X},\bm{Y})$,
\begin{align}
  \Pi = \Big\{P \in [0,1]^{n\times n}:\sum_iP_{ij}=\sum_jP_{ij}=1\Big\}.
\end{align}
The components $P_{ij}$ correspond to the fraction of $\bm{X}_i$ atoms mapped to $\bm{Y}_j$ sites.
The constraints $\sum_iP_{ij}=1 \ \forall j$ and $\sum_jP_{ij}=1 \ \forall i$ are required for to satisfy surjectivity and injectivity of $\mathcal{X}\mapsto\mathcal{Y}$, respectively.
(The matrices $P\in\Pi$ are doubly stochastic and can be parameterized by $(n-1)^2$ independent variables.
However the volume $|\Pi|$, equal to the volume of the Birkhoff polytope $B_n$, is not known in general for arbitrary $n$.)
The use of probability matrices $P$ instead of permutations $\sigma$ allow the cost function $C$ to be replaced with its expected value, and the program is expressed as
\begin{align}\label{eq:mongekantorovich_mod}
    &\inf_{P\in\Pi}\sum_{ij}C_{ij}P_{ij}
    & C_{ij} &= C(X_i,x_{j}).
\end{align}
Equation \ref{eq:mongekantorovich_mod} is a constrained linear program over the polytope $\Pi$, each corner of which corresponds to a permutation $\sigma\in\Sigma$.
In the present context, this program is somewhat over-constrained: given two permutations $\sigma_1,\sigma_2$ with nearly equal action, the algorithm will select only one of them.
In reality, such a degeneracy would likely produce a mixure of both permutations according to a probability determined by $P$.
That is, we expect in many cases solution for $P$ to lie in the interior of $\Pi$ rather than at the extreme points.
Towards this end, we apply entropic regularization.
This is an interior-point method that improves solvability by the addition of a convex entropic term.
The original discovery and use of the entropic regularization for transport problems has existed for some time and has been applied to a variety of transport problems, including automobile traffic and economics \cite{wilson1969use,erlander1979optimal,erlander1990gravity,svenaeus2012optimal}. 
The reader is referred to chapter 4 of \cite{peyre2019computational} for a comprehensive review of the subject.
The entropy of a bistochastic matrix is defined to be
\begin{align}
  H(P) = -\sum_{i,j}P_{ij}(\ln(P_{ij})-1).
\end{align}
Adding this to the objective function results in an unconstrained optimization problem,
\begin{align}\label{eq:mongekantorivich_mod_entropy}
    \inf_P\Big[ \sum_{i,j} \Big(C(\bm{X}_i,\bm{x}_j)\Big)P_{ij} + \varepsilon\,H(P)\Big],
\end{align}
where $\varepsilon$ is a numerical regularization parameter.

The min-shuffle hypothesis is stated precisely by Equation \ref{eq:mongekantorivich_mod_entropy} in the limit of small $\varepsilon$.
Equation (\ref{eq:mongekantorivich_mod_entropy}) is solvable via Sinkhorn's algorithm, which is described in \cite{peyre2019computational} and \cite{sinkhorn1964relationship}.
One immediately notices a strong similarity between (\ref{eq:mongekantorivich_mod_entropy}) and the maximum entropy formulation for thermodynamics, where $\varepsilon$ corresponds to $k_BT$.
It will be shown by means of multiple examples that $\varepsilon$ serves as a useful proxy for temperature when examining high temperature grain boundary migration.
However, we emphasize that any connection between $\varepsilon$ and $k_BT$ is heuristic rather than derived.
We urge caution in ascribing any principles of equilibrium thermodynamics to boundary migration, which is a decidedly non-equilibrium process.

Up to this point a constant value for $\U$ has been assumed; that is, $\bm{X}$ and $\bm{x}$ were assumed to be predetermined.
However, the transport problem makes no constraint on the microscopic translation vector, only on the crystallographic configuration before and after boundary motion.
In other words, while $\bm{X}$ and $\bm{Y}$ are fixed, $\U$ is not necessarily known.
Repeating the prior steps allowing for variable shifts $\U$, we have
\begin{align}\label{eq:mongekantorivich_mod_entropy2}
    \inf_{\U} \inf_P\Big[ \sum_{i,j} \Big(C(\bm{X}_i,\bm{Y}_j+\U)\Big)P_{ij} + \varepsilon\,H(P)\Big].
\end{align}
Stationarity for $\U$ produces the simple result for the mobility-optimal shift
\begin{align}\label{eq:solution_for_u}
  \U^*_M  = \frac{1}{n}\sum_{i,j} (\bm{X}_i - \bm{Y}_j) P_{ij}
\end{align}
where $n = |\bm{X}|=|\bm{Y}|=|\bm{x}|$ the number of atoms in the unit cell.
The optimal $P$ and $\U$ can be determined by an iterative application of Sinkhorn's algorithm and (\ref{eq:solution_for_u}), respectively.

\section{Methodology}\label{sec:methodology}

\subsection{Optimal transport computations}

Input point sets $\bm{X}$ and $\bm{Y}$ are sublattices of the dichromatic pattern which satisfy periodic boundary conditions for a predetermined CSL unit cell. In order to enumerate shuffle displacements for transformations with $\bm{F_{GB}} \ne \bm{I}$ corresponding to shear coupling, the net shear deformation is subtracted symmetrically from $\bm{X}$ and $\bm{Y}$. The Shuffling Dichromatic Pattern SDP($\bm{F_{GB}}$) is defined for each allowed choice of compatible shear transformation, similar to analysis in \cite{hirth2019topological,disptex}. 
In order to directly compare atomic trajectories from MD to predictions of the forward model, the microscopic translation vector $\U$ is precomputed from relaxed bicrystal structures in MD simulations. $\U$ is measured relative to the coherent dichromatic pattern \cite{disptex}. On the other hand, the distance minimizing (mobility-optimal) shift, $\U^*_M$ can be solved directly from Eqn \ref{eq:solution_for_u} without input from MD data. 

As shown in \cite[Proposition 4.3]{peyre2019computational}, Equation \ref{eq:mongekantorivich_mod_entropy} has a unique solution for fixed $\U$ given by

\begin{align}\label{eq:soln_kantorovich}
P_{ij} = \bm{f} \ e^{-C(\bm{X}_i,\bm{Y}_j+\U)/ \varepsilon} \ \bm{g} ,
\end{align}

where  $\bm{f}$ and $\bm{g}$ are scaling vectors that can be determined efficiently using Sinkhorn's algorithm \cite{altschuler2017near}. A MATLAB implementation of Sinkhorn's algorithm \cite{peyre2019computational} is used in this work with a cost matrix $C$ consisting of pairwise squared shuffling distances in the dichromatic pattern. 

\subsection{Molecular dynamics simulations}\label{sec:MD}

The ramped synthetic driving force method (rSDF) is used to generate grain boundary motion in bicrystal structures by applying a potential energy jump across the GB \cite{deng2017size}, biasing the growth of the low energy grain. The rSDF method extracts a critical driving force corresponding to the onset of motion in each GB.  Critical driving force values vary from experimentally accessible driving force magnitudes (less than 10 MPa) for highly mobile GBs to high values (on the order of GPa) for low mobility GBs \cite{chesserout}. Critical driving force values generally decrease with temperature, as described in \cite{deng2017size}. The rSDF method is convenient for examining GB migration over a large temperature range. Low temperature migration trajectories with minimal thermal noise establish an important baseline for displacement analysis as described in Section \ref{sec:dispseg}.  

The Janssens implementation of the synthetic driving force was used in this work \cite{Janssens2006}. The Janssens driving force has been shown to give consistent migration mechanism results compared to other techniques, including strain driven motion for equivalent mode selection \cite{coleman2014effect}, the random walk method \cite{Trautt2006} and the ECO implementation of the synthetic driving force \cite{ulomek2015enfJergy}. In the case of the Janssens driving force, the actual driving force applied to the bicrystal is often less than the value specified in LAMMPS because of noise in the order parameter at high temperatures and/or small disorientations. The actual driving force is computed directly from dump files via thermodynamic integration as specified by Olmsted in \cite{olmsted2009survey}.  At the time of writing, a new, more computationally efficient version of the ECO driving force has been published with $O(n)$ force calculations which avoids the need for thermodynamic integration \cite{schratt2020}. Consistency of the results from this work with the new ECO method will be left to future work.

\subsection{Displacement segmentation from MD simulations}\label{sec:dispseg}

Optimal shuffle displacements are enumerated from the forward model at small levels of regularization $\epsilon$. Non-optimal displacements may also be enumerated at larger values of $\epsilon$. Optimal shuffle displacements and, optionally, non-optimal displacements serve as a template for identifying the displacement types observed during MD simulations. Each displacement from MD data is compared to the template and either a close match is found or the displacement is classified as "other". The matching criterion involves comparison of displacement length and orientation in the MD versus template data. Note that the template and the MD data can be expressed in different reference frames, as described in \cite{disptex}. In this work, we compare the MD and OT data in the Optimal Translated Dichromatic Pattern (OTDP), the reference frame obtained by subtracting the net displacement per atom from the overall displacement pattern (equivalent to choosing a microscopic translation vector $\U^*$). This subtraction process makes displacement matching more robust against thermal noise or small sliding events at high temperatures. 

\section{Results and Discussion}\label{sec:results_and_discussion}

In this section, we demonstrate how the forward model can be used to test the min-shuffle hypothesis against MD data for several GBs. We show that non-optimal displacements play an important role in high temperature, high driving force, or high $\Sigma$ migration of twist GBs, assessing the applicability of the min-shuffle hypothesis and OT model predictions in this context. Next, we consider highly mobility $\Sigma 3$ GB migration and argue that this case is close to satisfying the OT model assumptions. We discuss the important role of the microscopic translation vector in determining the boundary plane anisotropy of shuffling patterns for $\Sigma 3$ GBs. Finally, we demonstrate via analysis of shear coupled migration of a symmetric tilt GB that optimal shuffles may be predicted for arbitrary disconnection modes that closely match MD data. 

\subsection{Testing the min-shuffle hypothesis for twist GB migration}

\begin{figure}[h]
    \centering\leavevmode
    \includegraphics[width=0.9\textwidth]{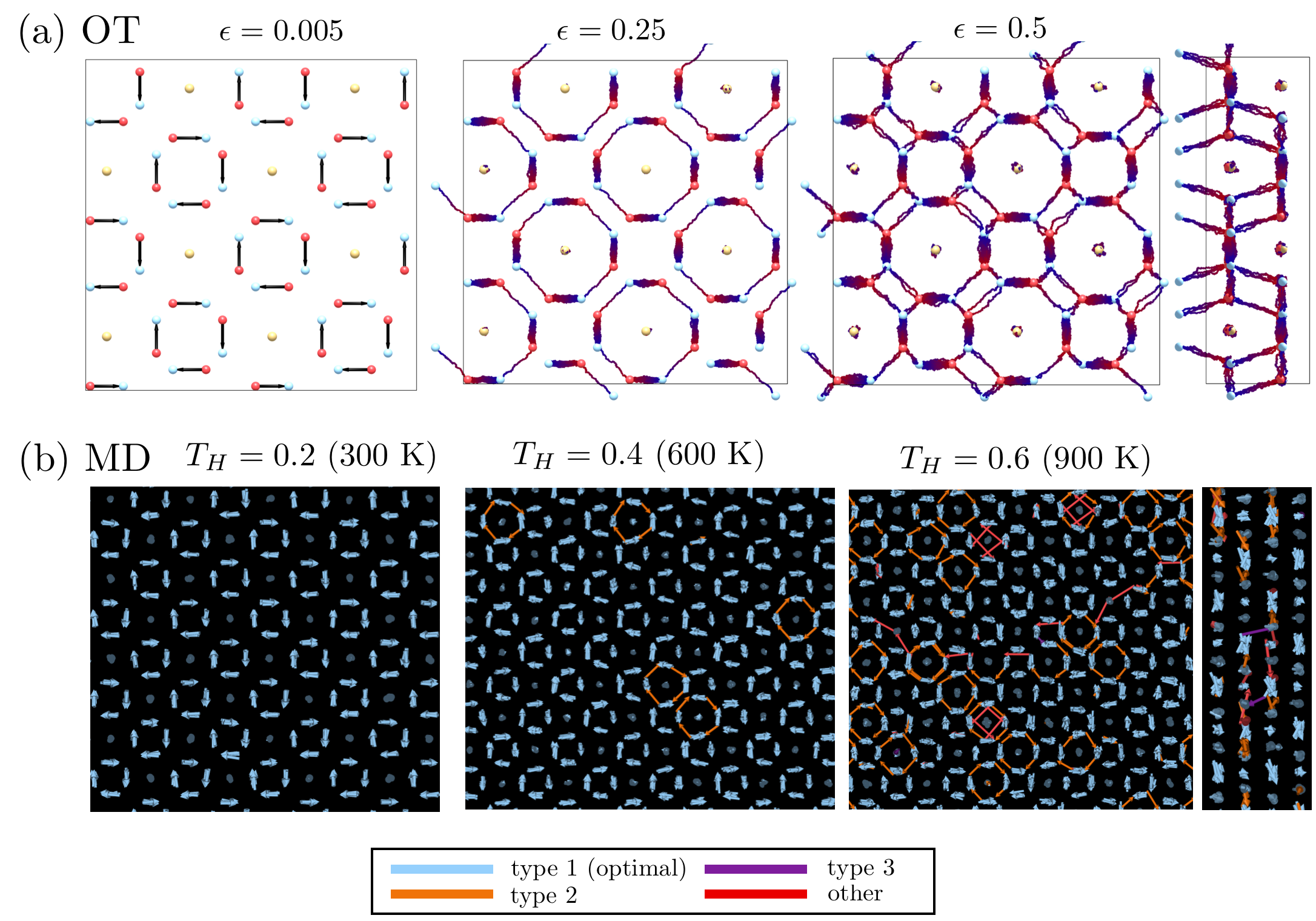}
    \caption{The diversity of non-optimal displacements increases with $\epsilon$ in the forward model and temperature in the MD data. (a) $\hkl{100}$ plane view of displacements predicted for the $\Sigma5 \hkl{100}$ GB. On far left, at small $\epsilon$, min-shuffle pattern is obtained (Type 1 displacements). In center ($\epsilon = 0.25$), non-optimal displacements are observed with nonzero probability (Type 2 displacements). The middle and rightmost images employ a brownian bridge visualization method \cite{peyre2019computational} which simulates a random walk between fixed initial and final lattice points with line density proportional to path probability. In rightmost two images, additional Type 3 displacements are observed ($\epsilon = 0.5$) between $\hkl{100}$ planes (rightmost image). (b) MD snapshots of GB migration at various temperatures viewed along $\hkl{100}$ plane. Displacements are colored by type with "other" classification (red) corresponding to any displacement not classified as Type 1-3.}
    \label{fig:twistprobs}
\end{figure}

Migration of the $\Sigma 5 \hkl{100}$ 36.87 $^\circ$ twist GB offers a clear example of ordered atomic displacements during GB migration, as shown in Figure \ref{fig:twistprobs}. Over a wide range of temperatures and driving forces, the min-shuffle transformation is found to be the most probable transformation. Collectively, min-shuffle displacements are labeled Type 1 displacements (blue in Figure \ref{fig:twistprobs}). The geometry of Type 1 displacements is consistent with prior MD studies which highlight a four shuffle motion mechanism during GB migration driven by an elastic strain or curvature driving force \cite{yan2010atomistic,bristowe}. The curvature driven migration study of the $\Sigma 5 \hkl{100}$ twist GB by Jhan and Bristowe was among the earliest MD simulations of GB migration \cite{bristowe}, motivated by the fast, jerky migration of near twist GBs in the in situ TEM experiments of Balluffi and Babcock \cite{babcock1989grain}. In \cite{babcock1989grain}, it was hypothesized that GB motion primarily occurs by collective atomic shuffles. Type 1 displacements are $\hkl<310>$ type shuffles which form loops around coincident sites in the dichromatic pattern (consistent with the CSL pinned migration mechanisms of $\Sigma 7$ and $\Sigma 9$ GBs in \cite{bair2019antithermal}). For the Ni GB under study, a small microscopic translation $\U$ normal to the twist plane (chosen on the basis of GB energy minimization) leads to a small displacement at each coincident site and a small out of plane component of each $\hkl<310>$ type shuffle vector. 

\begin{figure}[t]
    \centering\leavevmode
    \includegraphics[width=0.9\textwidth]{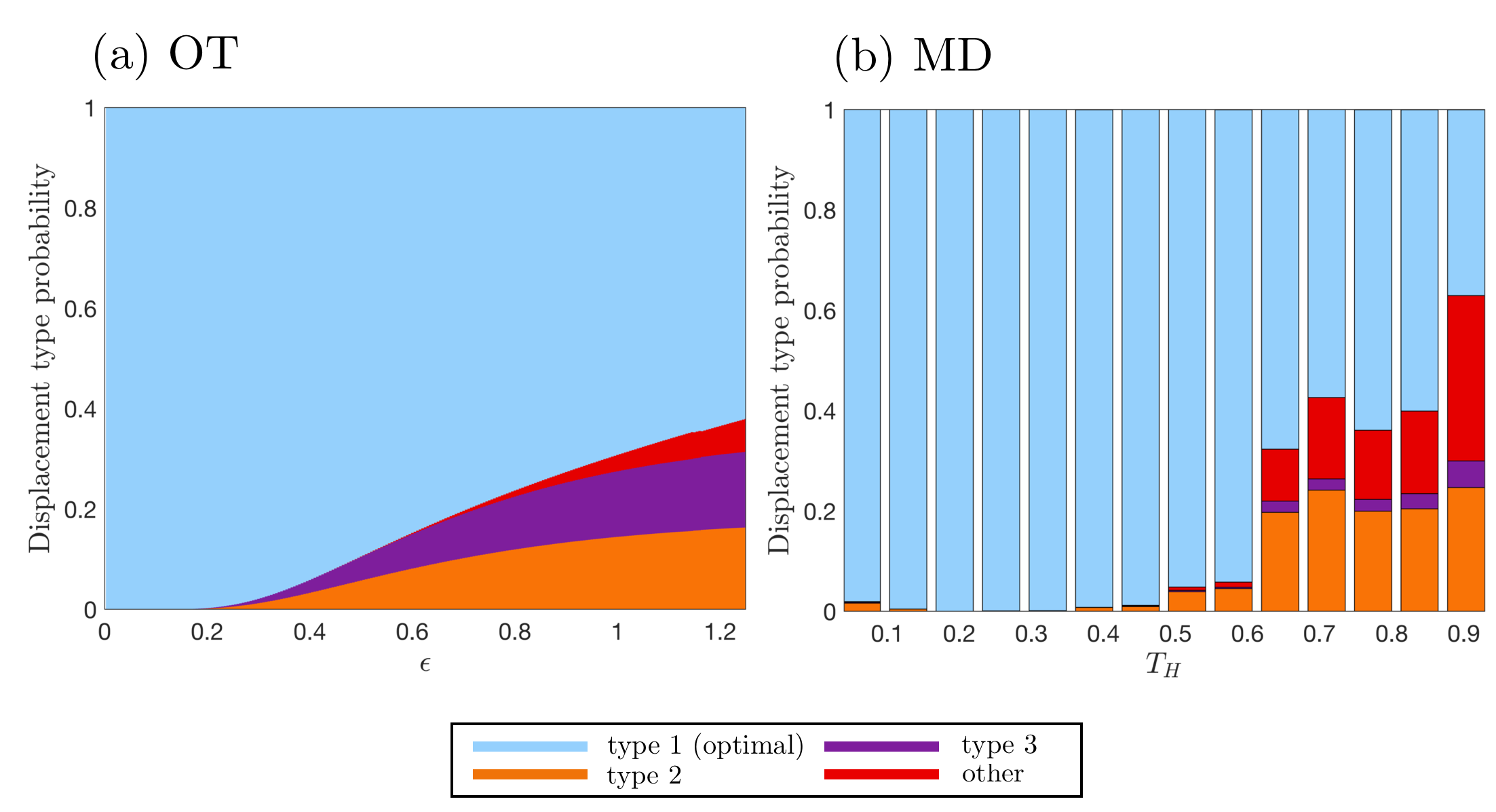}
    \caption{(a) Displacement type probabilities for energy driven migration of $\Sigma5$ $\hkl{100}$ GB under constrained boundary conditions (b) Displacement type probabilities from forward model are shown for the best fit $\epsilon$ range.}
    \label{fig:OTviz}
\end{figure}

The forward model allows classification of longer, non-optimal displacements which may compete with Type 1 displacements. For instance, Type 2 displacements (orange in Figure \ref{fig:twistprobs}) offer an alternative microrotation mechanism to Type 1 displacements with shuffles along $\hkl<210>$ directions. Compared to Type 1 displacements, Type 2 displacements have the second lowest net shuffling cost and have opposite handedness.  Type 2 displacement probabilities are nonzero at 100 K and decrease to near zero at 300 K as shown in Figure \ref{fig:OTviz}. This trend can be explained by the large critical driving force required for motion at low temperature (0.24 eV/atom at 100 K). The high applied driving force enables higher cost shuffles. The critical driving force decreases monotonically with temperature to near 1 meV/atom at 1400 K. At intermediate temperatures 300-500 K, the min-shuffle transformation dominates the statistics of migration to within 0.001 probability. As temperature is increased further, Figure \ref{fig:OTviz} shows that non-optimal displacements increase in probability. In addition to Type 2 displacements, Type 3 displacements (purple in Figure \ref{fig:twistprobs}) are observed between $\hkl{100}$ planes. All other displacements are classified as ``other'' and include large displacements for which multiple atomic jumps may have occurred (red in Figure \ref{fig:twistprobs}). At high temperatures above 1000 K, the probabilities of higher order displacements remain approximately constant with the probability of "other" displacements increasing at 1400 K. These "other" displacements appear to be associated with GB self diffusion and sliding events during migration. In \cite{yan2010atomistic}, so called stringlike displacements of the $\Sigma 5 \hkl{100}$ GB in Ni were analyzed at 900 K under an elastic strain driving force \cite{yan2010atomistic}. Stringlike displacements are a type of dynamic heterogeneity common to diverse physical phenomena including glass formation, nanoparticle melting, and GB diffusion \cite{Lacevic2004,dauchot2005dynamical,mishin2015atomistic}. During stringlike motion, mobile atoms follow each other in a train-like sequence of hops called a string with a characteristic probability that depends on string length and only weakly on temperature \cite{mishin2015atomistic}. Stringlike displacements can occur simultaneously with GB migration (with or without shear coupling) \cite{zhang2006characterization,zhang2007atomic}. In \cite{yan2010atomistic}, a correlation was found between average string length and activation energy for three $\hkl{100}$ twist GBs. Since long displacements that are the net result of many atomic jumps violate the assumption of geodesic closeness, stringlike displacements are out of the scope of the current OT model. Nevertheless, the forward model provides a geometric interpolation scheme to classify different types of shuffle displacements from MD data, including net displacements resulting from stringlike motion.


\begin{table}[H]
\setlength\tabcolsep{0pt} 
\caption{Parameters for min-shuffle transformation. ($\ast$) from MD data at 100 K}
\label{tab:ms}
\begin{tabular*}{\textwidth}{@{\extracolsep{\fill}} l *{5}{d{2.4}} }
\toprule
 & \multicolumn{1}{c}{$n$} & \multicolumn{1}{c}{$d^2$ ($\AA^2$)} & \multicolumn{1}{c}{U (eV/atom)} & \multicolumn{1}{c}{$\tau$ (ps)}  & \multicolumn{1}{c}{$\mathcal{D}$} \\
\midrule
$\Sigma 5$\hkl{100} & 10 & 10.96 & 0.27 \ast &  0.8\ast & 51.8 \\

$\Sigma 3$\hkl{110}  & 6   & 4.94 & 0.0018 \ast & 0.8\ast & 0.46 \\

$\Sigma 85$\hkl{100} & 170 & 188.55 & 0.29 \ast &  <2.5\ast & 537 \\
   
\bottomrule
\end{tabular*}
\end{table}

 \begin{figure}[t]
    \centering\leavevmode
    \includegraphics[width=0.9\textwidth]{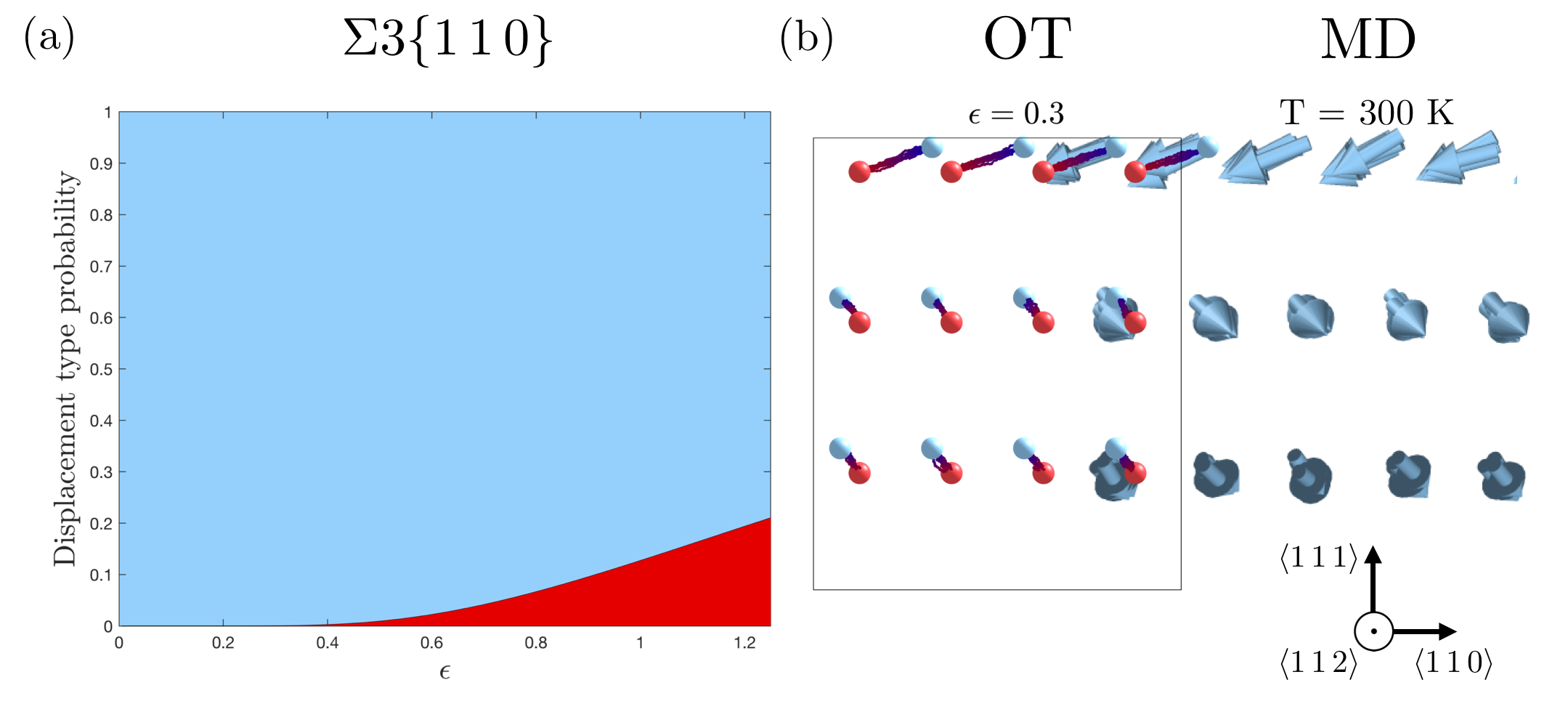}
    \caption{(a) Displacement type probability predictions from forward model for $\Sigma3\hkl{110}$ GB. Color scheme is consistent with Figure \ref{fig:twistprobs}. (b) Displacement pattern prediction from forward model is overlaid on displacement pattern from MD data at 200 K (blue displacements). A 3D perspective of this displacement pattern is shown in Figure \ref{fig:pspace}. In the MD data, the probability of optimal displacements is large ($> 99.99\%$) under all simulated conditions.}
    \label{fig:twist_comps}
\end{figure}


As shown in Figure \ref{fig:OTviz}, the forward model achieves surprising agreement with the displacement type probability data tabulated from MD simulations given its simple assumption of distance minimization. In the case of the $\Sigma 5 \hkl{100}$ twist GB, model assumptions of small $\mathcal{D}$ are in fact \textit{not} satisfied because the potential energy barrier for migration is large relative to the kinetic energy term (see Table \ref{tab:ms}). Nevertheless, a positive correlation is found between the net transformation distance and energy barrier for GB migration computed by the Nudged Elastic Band (NEB) method for several trajectories from the forward model (\ref{sec:supplementary_information}, Figure~\ref{fig:S1}). Type 1 mappings (smallest net distance) have the lowest energy barrier, followed by Type 2 mappings and mappings which combine Type 1-3 displacements. Note that these computations do not explicitly include relaxed GB structure apart from the precomputed shift in the dichromatic pattern. Several aspects of displacement statistics from the OT model (which assumes small $\mathcal{D}$) lend interpretability to the MD data. In the forward model, non-optimal displacement type probabilities increase with regularization parameter $\epsilon$. The model captures the correct ordering of displacement probabilities ($P(1) > P(2) > P(3)$) observed in the MD data at low and intermediate temperatures for small and intermediate values of $\epsilon$. The sigmoidal shape of the Type 2 displacement probability-temperature curve from the MD data is reproduced by the forward model. However, the relative magnitudes of the probabilities predicted by the OT model do not match the MD data. Type 3 displacement probabilities are systematically overestimated and higher order displacements classified as "other" cannot straightforwardly be compared to MD data with multiple jumps. Ultimately, a best fit $\epsilon$ scaling factor is determined that minimizes the sum of the root mean square error between MD and OT displacement type probabilities. The $\epsilon$ scale 0-1.25 shown in Figure \ref{fig:twist_comps}a is found to be the best fit with an $R^2$ value of 0.74. It should be noted that this best fit is computed using temperature as a proxy for $\epsilon$. Other possible proxies including combination of temperature and critical driving force were found to give poorer fits. In summary, moderate agreement is found between the MD data and forward model despite violation of the small $\mathcal{D}$ assumption. 

Next, we consider the motion of the highly mobile $\Sigma 3$ \hkl(110) twist GB and argue that this GB is close to satisfying the optimal transport model assumptions ($\mathcal{D} < 1$). Prior MD calculations for this GB indicate a very small energy barrier to motion associated with partial dislocation glide \cite{Humberson2017,Humberson2019,Chesser2018}. General $\Sigma 3$ GBs that migrate via \hkl{110} facet motion have mobilities orders of magnitude larger than other $\Sigma 3$ GBs and among the highest mobilities in the 388 FCC Ni GB mobility survey \cite{olmsted2009survey}. For these GBs, mobility attains a maximum at cryogenic temperatures and decreases with increasing temperature, similar to phonon damped dislocation glide \cite{Humberson2017,Chesser2018,Cantwell2015}. In this work, the $\Sigma 3$ \hkl(110) GB is found to migrate via the min-shuffle transformation under all simulation conditions tested (free and fixed boundary conditions, 100-1400 K) and has  $\mathcal{D} < 1$ (Table \ref{tab:ms}). The characteristic relaxation time associated with one unit cell of displacements from start to finish is approximately 0.8 ps at 100 K with only small deviation from linear trajectories. As shown in Figure \ref{fig:twist_comps}b, the min-shuffle migration mechanism is characterized by three shuffling vectors along two $\hkl<112>$ and one $\hkl<110>$ type direction (with magnitude smaller than a full dislocation), consistent with a picture of boundary migration mediated by triplets of partial dislocations \cite{Humberson2019}. The distribution of each shuffling vector orientation broadens with temperature and very few higher order displacements are resolved outside of the thermal noise profile (less than 0.1 percent of displacements). Prior slip vector analysis for the $(110)$ ICT in \cite{priedeman2017role} originally reported a mechanism with nine displacement types, but this analysis was later amended to a three shuffle mechanism in \cite{bair2019antithermal} upon more careful analysis of thermal noise. In summary, the fast three shuffle mechanism of the $\Sigma 3$ \hkl(110) twist GB observed in this work is found to be optimal under all test conditions. 

Predictions of the forward model for the $\Sigma 3$ \hkl(110) twist GB are shown in Figure~\ref{fig:twist_comps}a. At equivalent regularization values, fewer higher order displacements are predicted for the $\Sigma 3$ GB than the $\Sigma 5$ GB, consistent with the MD data. The best-fit $\epsilon$ scaling factor for the $\Sigma 3$ GB is much smaller (0.3) than for the $\Sigma 5$ GB (1.25), reflecting the fact that optimal displacements strongly dominate the reorientation geometry in the MD data.

 \begin{figure}[t]
    \centering\leavevmode
    \includegraphics[width=0.8\textwidth]{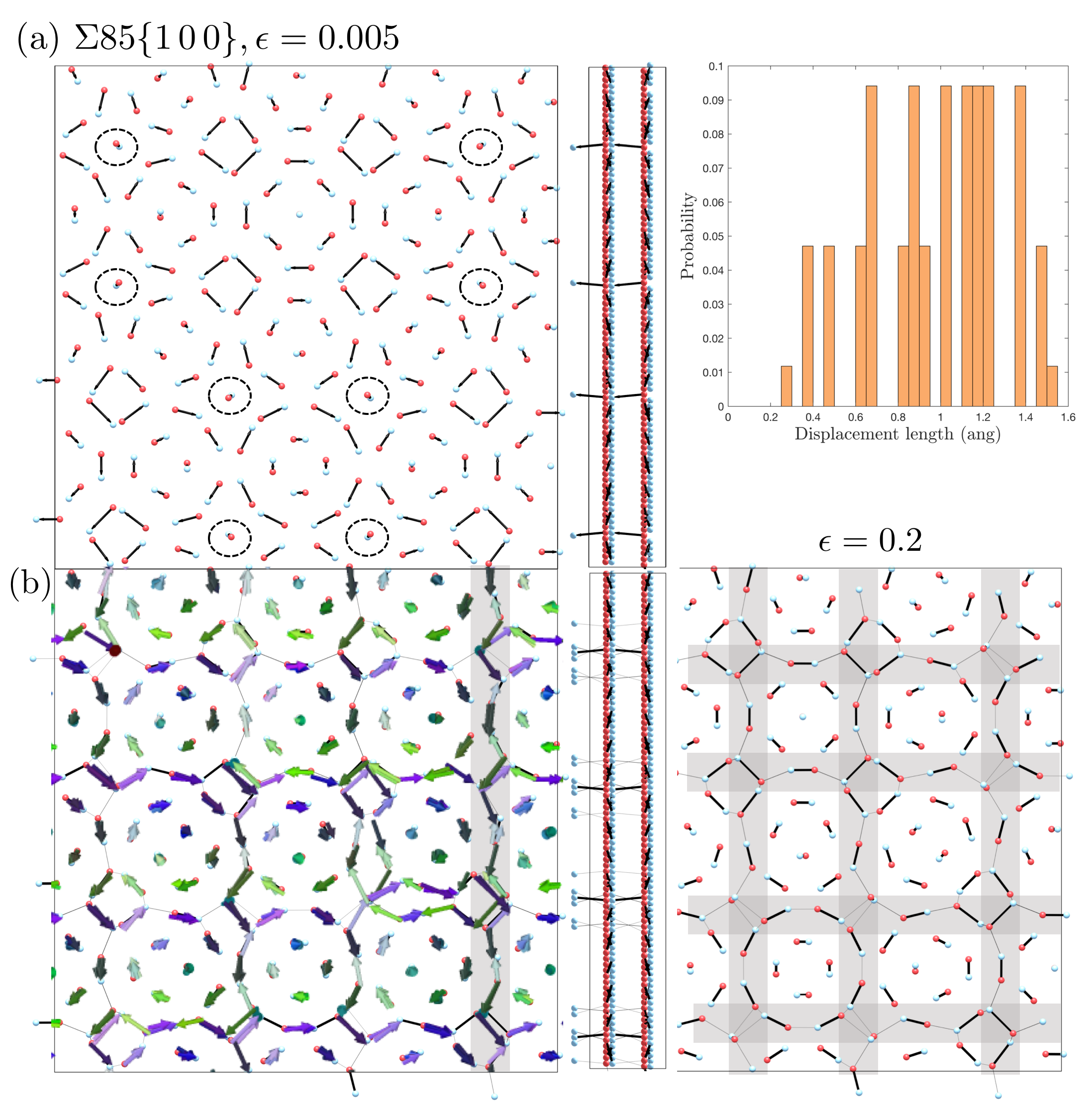}
    \caption{(a) The min shuffle pattern ($\varepsilon = 0.005$) for the $\Sigma 85$ $\hkl{100}$ twist GB (LHS, middle) has a variety of displacement lengths (RHS), including six relatively large displacements normal to the $\hkl{100}$ GB plane (circled in plane view on LHS, shown in side view in middle panel). (b) Direct comparison of MD data at 600 K (LHS, middle) to predicted displacement texture at $\varepsilon = 0.2$ (RHS) reveals that all minimal shuffle displacements can be found in the MD data. Additional displacements associated with bands of higher order displacements (highlighted in grey) are observed in both the MD data and the forward model for sufficiently large regularization. The localization of higher order displacements resemble a network of screw dislocations.}
    \label{fig:hisig1}
\end{figure}

Now we consider the motion of the  $\Sigma 85 \hkl{100}$ 25.06 $^\circ$ twist GB as shown in Figure \ref{fig:hisig1} (id 252 in \cite{olmsted2009survey}). Even though $\mathcal{D}$ is large, the forward model offers useful geometric insight in this complex setting. Since high $\Sigma$ GBs have large CSL unit cells, many types of shuffles are required to accommodate motion, even in the min-shuffle transformation. For instance, out of plane displacements between $\hkl{100}$ planes are required in the min-shuffle transformation for the $\Sigma 85$ GB, in contrast to the $\Sigma 5$ GB. Visual comparison of the MD migration data for the $\Sigma 85$ GB to the min-shuffle pattern from the forward model indicates that many of the displacements in the MD data are optimal. However, non-optimal displacements are always observed in the MD data in regions where relatively long shuffles are required, even at low and intermediate temperatures. These non-optimal displacements form bands in the plane of the GB (highlighted in grey in Figure \ref{fig:hisig1}b) which are captured by regularization in the forward model. In addition, the number and diversity of plane normal displacements increases with regularization in the forward model and with temperature and driving force magnitude in the MD data. In \cite{yan2010atomistic}, banded displacement patterns for a $\Sigma 13$ twist GB were interpreted as a network of stringlike displacements within screw dislocation cores in the plane of the GB. 

The relatively large characteristic time scale and critical driving force of migration of the $\Sigma 85 \hkl{100}$ GB in MD simulations leads to a large value of $\mathcal{D}$ (Table \ref{tab:ms}). 
When $\mathcal{D}$ is small, the reduction to (\ref{eq:wasserstein}) is exact because the potential effects are negligible.
However, when $\mathcal{D}$ becomes large, the unique effects of the interatomic potential dominate over the kinetic effects, and the result will be entirely potential-dependent.
In the MD simulations, the transformation proceeds in stages, starting with ringlike migration of 13 atom clusters along a single $\hkl{100}$ plane. This 13 atom activation volume is relatively small compared to the entire CSL unit cell (170 atoms) and initiates in regions which later become bands of higher order displacements. From the disconnection perspective, these clusters of atoms are nuclei with pure step character. The forward model captures the geometric combination of shuffles associated with step nucleation and stringlike diffusion, but does not give direct insight into nucleation and growth kinetics or rate limiting reaction steps. 

In all three twist GBs studied, the probability of optimal displacements exceed the probability of non-optimal displacements at temperatures 100-1000 K. This implies that, for driven grain boundary migration of these GBs at low to intermediate temperatures, the min-shuffle transformation dominates the geometry of reorientation, offering support for the min-shuffle hypothesis. In the case of the $\Sigma 5$ and $\Sigma 85$ twist GBs, $\mathcal{D}$ is large and non-optimal displacements are observed which are constrained by bicrystallography and resemble GB diffusion. These non-optimal displacements become more probable with increasing temperature or driving force, can dominate the character of migration at sufficiently high temperatures, and are expected to be interatomic potential dependent. Whether these displacements participate in rate limiting reaction steps during GB migration is an important question beyond the scope of the current work. On the other hand, the migration of the highly mobile $\Sigma 3$ GB is entirely dominated by the min-shuffle transformation. In this case, ordered shuffling can be considered a rate limiting mechanism and is predicted to resemble a second order phase transition with near uniform migration without the need for nucleation \cite{cahn1960theory}. 
It should be noted that while the OT based model captures the geometry of the transformation, it does  not include the thermodynamic driving force.
Additionally, the resulting solution has artificially low entropy, corresponding to an essentially ``boundary-free'' transformation in which the entire bicrystal transitions at once. A thermodynamic treatment of low  $\mathcal{D}$, high mobility migration is an interesting direction for future work.

\subsection{The impact of metastable states on displacement geometry}

An important geometric consequence of the OT model is that the min-shuffle transformation depends on small translational shifts $\U$ in the dichromatic pattern known as microscopic translational degrees of freedom. Different values of $\U$ may result in distinct shuffling patterns. Grain boundary mobility is hypothesized to depend on metastable state via variation in the min-shuffle transformation. In atomistic GB simulations, $\U$ is typically chosen to minimize GB free energy. The optimal $\U^*$ which minimizes shuffling distance is not necessarily the same as the energy optimal  $\U$, illustrating a tension between low shuffling cost associated with high mobility and low GB energy. 

\begin{figure}[t]
    \centering\leavevmode
    \includegraphics[width=0.8\textwidth]{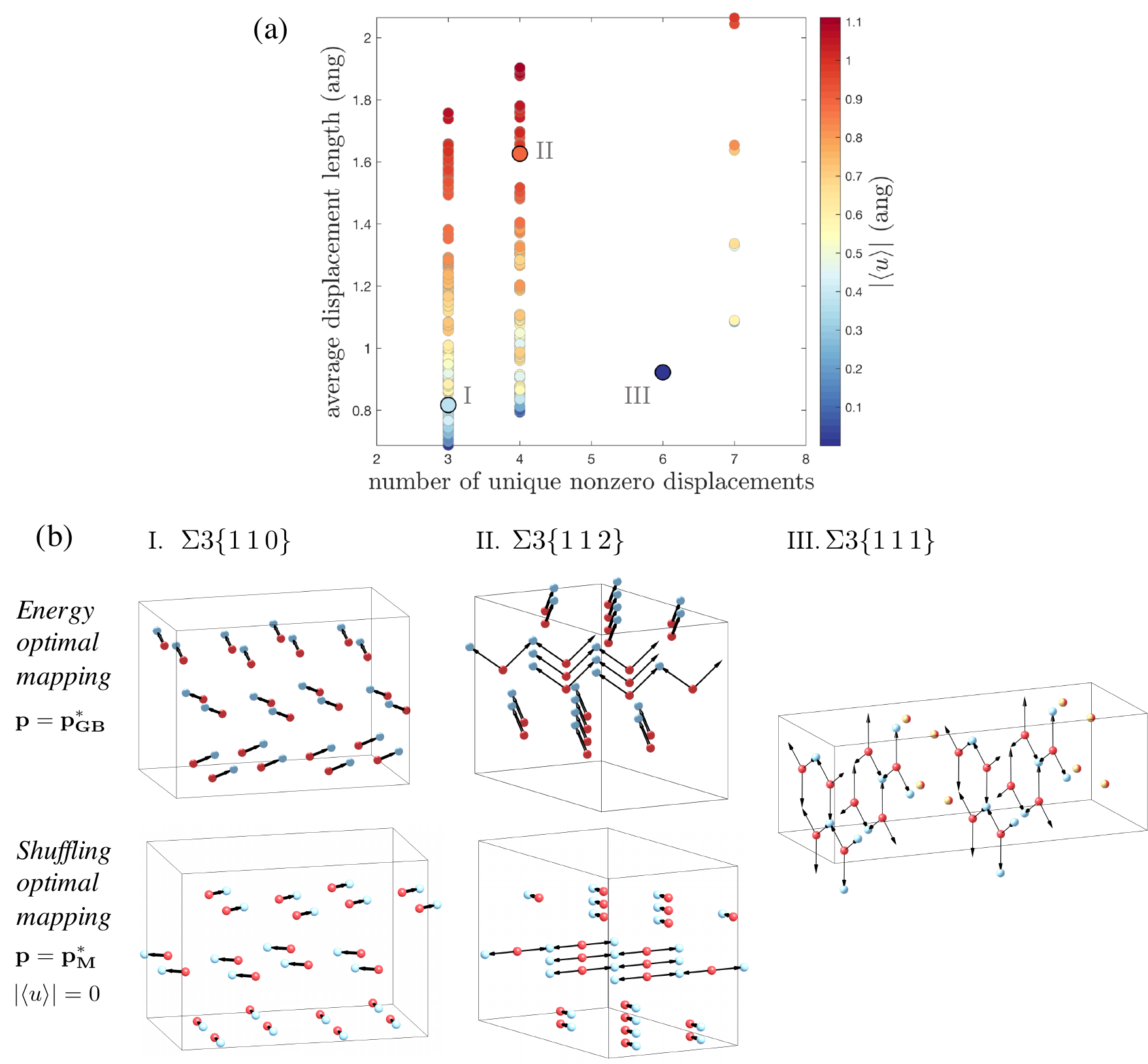}
    \caption{(a) Performing a grid search over allowed microscopic translations in the $\Sigma 3$ dichromatic pattern, four classes of min-shuffle transformations are found with 3, 4, 6 and 7 unique displacement vectors. Three of these transformations are observed in MD data for varying boundary plane inclinations. When the microscopic translation in these transformations is chosen on the basis of GB energy minimization, the shuffling cost of these transformations (circled in (a)) is not a minimum within the 3 and 4 displacement transformation classes, highlighting the difference between energy optimal and shuffling optimal transformations. The \hkl{110} GB (I) has 3 displacements, the \hkl{112} GB (II) has 4, and the {111} GB (III) has 6.}
    \label{fig:pspace}
\end{figure}

$\Sigma 3$ GBs provide an interesting example of the impact of $\U$ on displacement geometry. Surprisingly, the boundary plane anisotropy of shuffling patterns observed in MD simulations of $\Sigma 3$ GB migration may be explained by varying $\U$ for the min-shuffle prediction of the forward model. Depending on GB plane inclination, transformations with 3, 4 or 6 shuffles were observed during energy driven motion of $\Sigma 3$ GBs at low and intermediate temperatures \cite{disptex}. Using the forward model, we can enumerate all possible min-shuffle transformations for $\Sigma 3$ GBs via a grid search over possible $\U$ in the Displacement Shift Complete (DSC) unit cell (\cite{Sutton1996}). This analysis reveals four classes of transformations with 3, 4, 6 or 7 unique shuffling vectors respectively. The first three classes of transformations are consistent with the MD data. Min-shuffle transformations representative of these transformations are shown in Figure \ref{fig:pspace} with $\U$ precomputed to match the GB energy optimal shift vector for the three predominant $\Sigma 3$ facet orientations with \hkl{111}, \hkl{112} and \hkl{110} boundary plane inclinations. Importantly, the \hkl{112} and \hkl{111} GBs have multiple one to one mappings available with the same total displacement length. This length degeneracy is depicted by multiple vectors connecting initial atomic positions to final positions and explains the variation in the number of shuffling vectors for different transformations. In the MD data, the degenerate displacements are distributed in distinct ways throughout the volume transformed by grain boundary motion, often. For instance, during the motion of the coherent twin boundary with \hkl{111} GBs, sliding events are observed as the GB moves in which an entire plane of atoms displaces by one of three possible twinning partials with magnitude $\frac{a}{6}\hkl<112>$. Over the course of motion, the GB switches between all six types of  $\hkl<112>$ displacements. Further analysis of degenerate displacements is given in \cite{disptex}. The forward model provides a useful framework for enumerating variation in displacement patterns with respect to microscopic degrees of freedom and changes in boundary plane inclination. 

$\Sigma 3$ GBs also illustrate the tension between low shuffling cost and low grain boundary energy. Globally, shuffling cost is minimized for $\Sigma 3$ GBs via a 3 shuffle transformation with  $\U^* = X$. Similarly, within each transformation class characterized by the number of unique displacement vectors, there exists a minimum cost with respect to $\U$. The $\U$ precomputed on the basis of grain boundary energy leads to shuffling costs higher than the minimum costs within each class (compare circled points in Figure \ref{fig:pspace}a to the points at the bottom of each band). This discrepancy illustrates that grain boundary energy optimal transformations do not necessarily minimize shuffling cost. The mobility of $\Sigma 3$ GBs differs between transformation classes with average mobility ordering $M_3 > M_4 > M_6$. It is hypothesized that within each class (along a band of costs), differing displacement lengths will also impact mobility by modifying the activation energy for shuffling. 


\subsection{Extension to shear coupled migration and multimodal data analysis}

In this section, the forward model is extended to predict shuffles during shear coupled motion with the advantage (compared to MD simulations) that the motion of arbitrary disconnection types may be considered. Up until this point, we have only considered GBs that sweep out zero net shear with negligible sliding character. In order to separate the sliding character of GB motion from the step character, we employ the shear-shuffle decomposition described in \cite{disptex}. In this framework, shear displacements that contribute to the net deformation are separated from shuffle displacements which do not. We proceed to compute the min-shuffle transformation for shear coupled migration of the $\Sigma 5$ $\hkl{310}$ GB with coupling factor $\beta_{110} = -1$. This transformation corresponds to the well known negative ${110}$ type coupling mode observed in simulations and experiments \cite{cahn2006a,molodov2018grain}. At low temperatures, the min-shuffle hypothesis is supported for this GB. On the other hand, at high temperatures we find a loss of shear coupling of this GB associated with many displacement types. We show that the mixture of displacement types at high temperatures is consistent with shuffles for two types of disconnections with opposite sign of shear. 

\begin{figure}[t]
    \centering\leavevmode
    \includegraphics[width=0.75\textwidth]{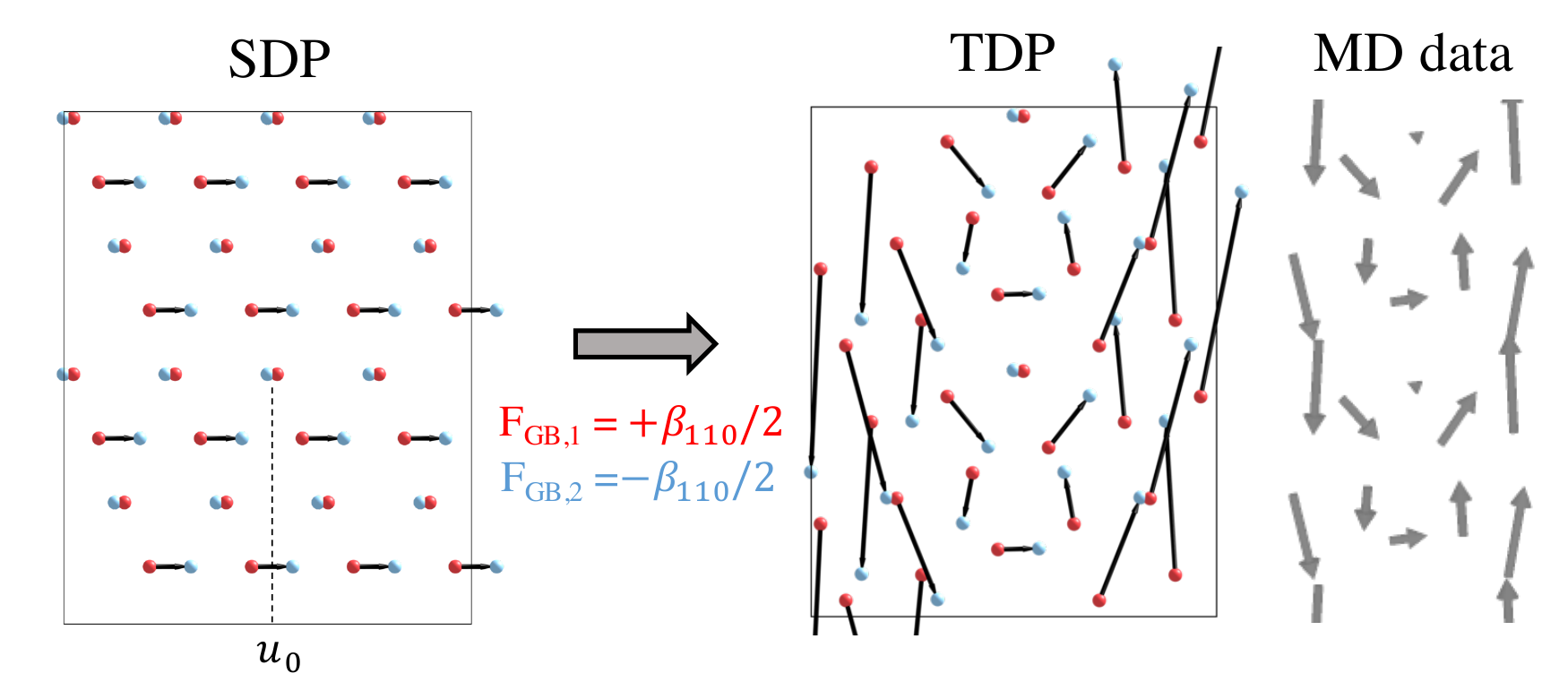}
    \caption{Shuffles for the $\hkl{110}$ mode of the $\Sigma 5$ $\hkl{310}$ GB enumerated in the SDP (deformed coordinates) and shown in the undeformed coordinates of the TDP match MD data well}
    \label{fig:onemode}
\end{figure}

The shuffling dichromatic pattern (SDP) \cite{hirth2019topological} is the appropriate dichromatic pattern representation for enumerating shuffles in the forward model. Construction of the SDP requires shear deformation consistent with a known coupling factor to be subtracted from the sublattices of the translated dichromatic pattern (TDP), as shown in Figure \ref{fig:onemode}. Note that the TDP is related to the coherent dichromatic pattern (CDP) by applying a relative shift $\U$ to the red sublattice in the CDP \cite{disptex}. There are several choices for how to subtract shear from the TDP, including subtracting shear from only one or both sublattices. We choose to construct the SDP by symmetrically subtracting shear deformations with components $\beta/2$ and $-\beta/2$ from each of the sublattices of the TDP. The regularized optimal transport problem is solved in the SDP to obtain the min-shuffle transformation in the limit of small  $\varepsilon$. In order to directly compare displacement patterns to raw MD data, the deformation that was originally subtracted from the dichromatic pattern is added back to the min-shuffle transformation symmetrically about a coordinate ${u_0}$. In Figure \ref{fig:onemode}, this process is depicted for shear coupled migration of the $\Sigma 5$ $\hkl{310}$ GB with coupling factor $\beta_{110} = -1$. Displacement predictions of the forward model are shown to match atomistic simulation data for strain driven migration of the $\Sigma 5$ $\hkl{310}$ GB at 0 K. Note that there are only two unique shuffling vectors in the SDP compared to the four observed for zero shear transformation of the $\Sigma 5$ twist GB. The shears during disconnection motion can reduce the number of unique shuffling vectors required during GB motion compared to the zero shear case. 

 \begin{figure}[t]
    \centering\leavevmode
    \includegraphics[width=0.7\textwidth]{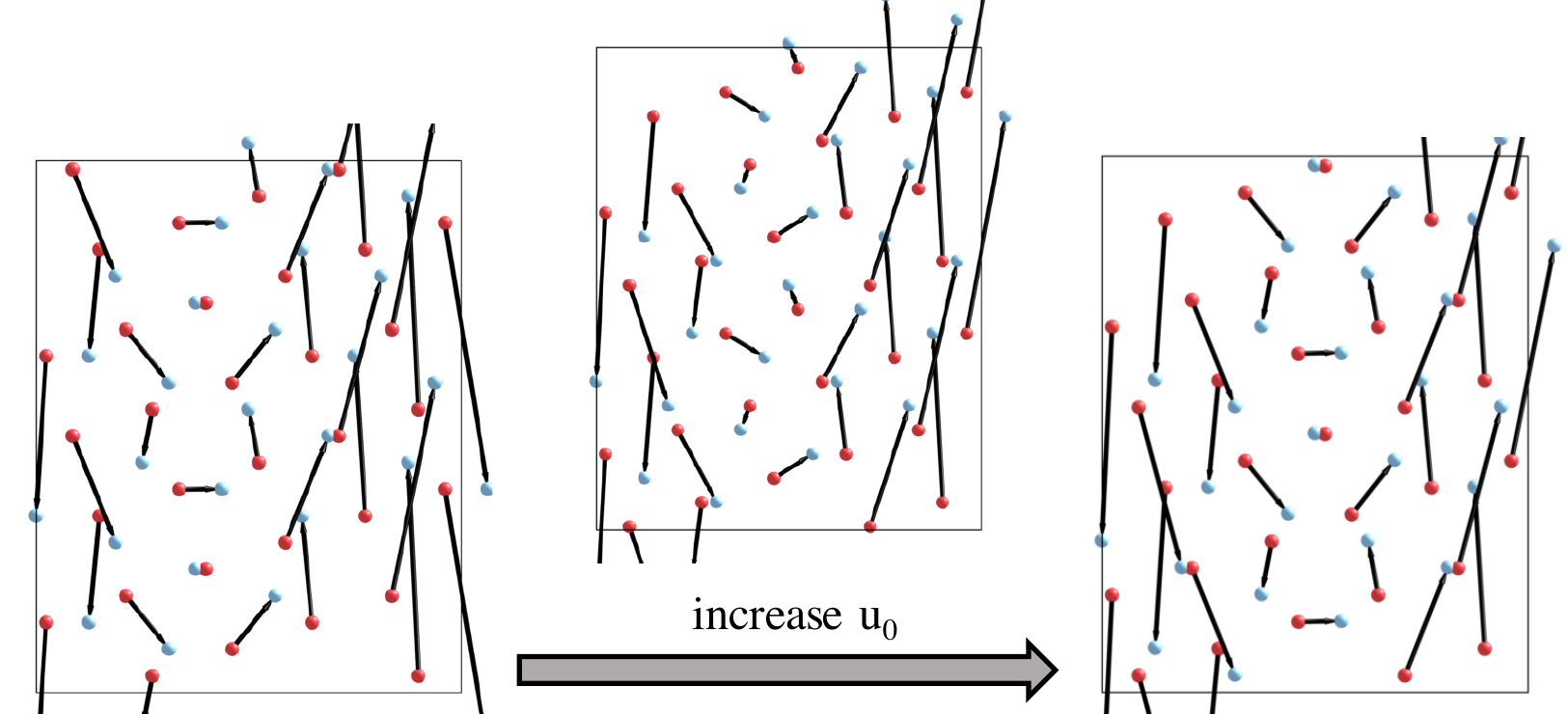}
    \caption{Different choices of $\bm{u_0}$ lead to equivalent displacement patterns in the undeformed coordinates of the TDP up to a rigid shift. This shift captures disconnection burgers vector in vertical direction and step height in lateral direction.}
    \label{fig:umode}
\end{figure}

 \begin{figure}[t]
    \centering\leavevmode
    \includegraphics[width=0.8\textwidth]{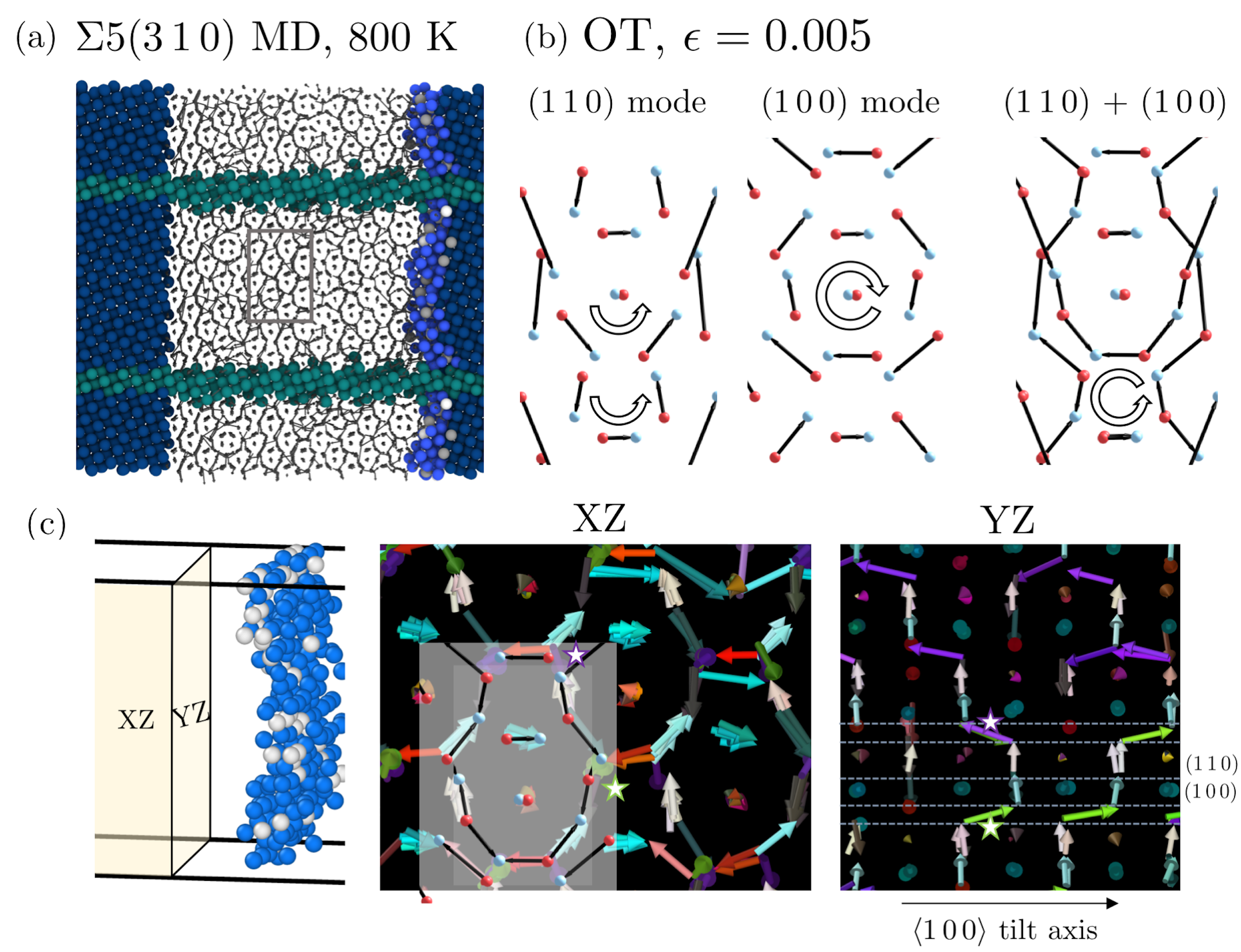}
    \caption{(a) MD snapshot of migration at 800 K under constrained boundary conditions for the $\Sigma5 \hkl{310} \hkl<100> 36.9^\circ$ GB. (b) shuffling pattern predictions of forward model in the undeformed frame for $\hkl{100}$ mode, $\hkl{110}$ mode and a combination of the two modes. (c) displacement texture viewed down the tilt axis (middle) and along GB plane (right) indicates that the motion mechanism is a mixture of $\hkl{100}$ and $\hkl{110}$ type displacements with network of non-optimal displacements along the tilt axis.}
    \label{fig:twomodes}
\end{figure}

 \begin{figure}[h]
    \centering\leavevmode
    \includegraphics[width=0.9\textwidth]{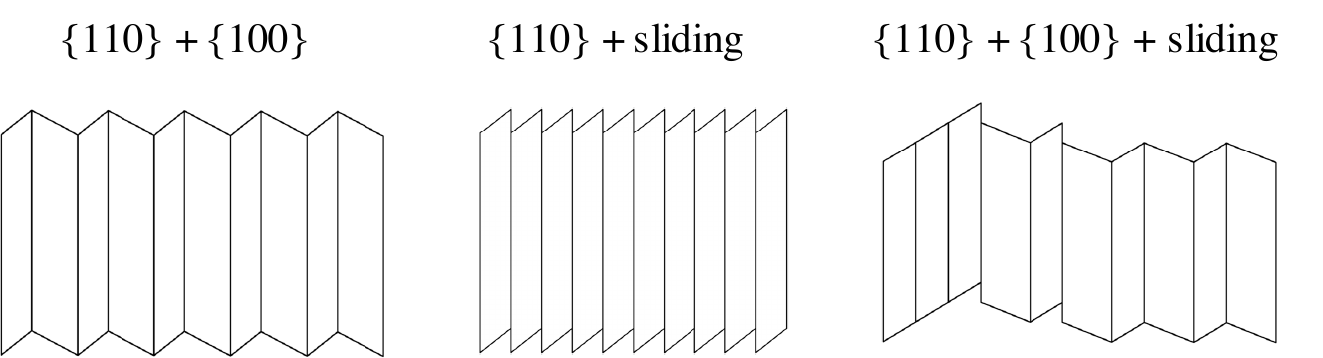}
    \caption{Several possible options for accommodating zero shear boundary conditions. Unlike the pictures above, most multimodal transformations observed in MD data require a 3-D picture involving atomic motion in the plane of the GB.}
    \label{fig:zigzag}
\end{figure}


Analysis of shear coupled migration in the forward model can also identify 1) different displacements arising from variations in microscopic translation vector 2) non-optimal displacements (frequently associated with GB diffusion) and 3) displacements corresponding to other disconnection modes. There is a connection between (1) and the definition of disconnections. The small rotation in the MD displacement data (in Figure \ref{fig:onemode}) relative to the prediction of the forward model in the TDP is a result of the choice of $\bm{u_0}$. A small offset of $\bm{u_0}$ to the left of the choice shown would lead to an asymmetric displacement pattern that matches MD data exactly. Disconnection step height $h$ may be inferred from the periodicity of the shuffling pattern with respect to choice of $u_0$ in the SDP.  This is shown pictorially for the $\hkl{110}$ mode in Figure \ref{fig:umode}.  The computed $h_{110} = 0.11$ nm is consistent with the analytical formulas for disconnection geometry of STGBs in Appendix A of \cite{han2018grain} using lattice parameter $a_{\text{Ni}} = 0.352$ nm. Different shuffling patterns with $u_{0,x} \in (0,h)$ could be observed for different metastable states of the $\Sigma 5$ $\hkl{310}$ GB with different out of plane components of the rigid shift. 

In this work, we demonstrate a relationship between non-optimal displacements and optimal displacements for multiple disconnection modes. We argue via detailed analysis of MD data that optimal displacements for two disconnection modes can mix at sufficiently high temperatures, but that this mixture is mediated by geometrically necessary higher order displacements associated with small sliding events. Our test case is intermediate and high temperature energy jump driven migration of the $\Sigma 5$ $\hkl{310}$ GB in which the shear coupling factor decreases to near zero under both free and fixed boundary conditions (Figure \ref{fig:twomodes}a). Zero shear migration of this GB was previously observed in simulations with periodic boundary conditions normal to the GB plane in \cite{schratt2020}. In  \cite{schratt2020}, multiple possible mechanisms were proposed including zig-zag motion via two alternating disconnection modes or single mode motion plus sliding. These mechanistic possibilities are summarized at the continuum level in Figure \ref{fig:zigzag}. We will show that these continuum pictures are insufficient to describe the complexity of the observed motion. 

Remarkably, when the min-shuffle patterns of two distinct disconnection types are superimposed for the $\Sigma 5$ $\hkl{310}$ GB, a displacement profile is generated with striking visual similarity to the displacements swept out by the zero shear migration of this GB in MD simulations. Predictions of the forward model for displacements during $\hkl{100}$ and $\hkl{110}$  type disconnection motion are shown in Figure \ref{fig:twomodes}b for small regularization ($\varepsilon = 0.005$). Displacements circulate with opposite handedness in the two modes, reflecting the different signs of shear. Although all of the shuffles from the two disconnection types can be found in the MD data, additional non-optimal displacements are observed. For instance, the long light blue displacements and purple and green displacements along the tilt axis (into the page) are observed in the MD data and are not observed in the low $\epsilon$ predictions of the forward model. Regularization can capture these displacements at sufficiently high regularization, but the mixture of non-optimal displacements at such high $\epsilon$ values does not match the non-optimal displacements in the MD data. 

Viewing the displacement pattern from the MD data in the plane of the GB (Figure \ref{fig:twomodes}c, right) clarifies the mixture of displacements during migration. Progressing upward along the $\hat{z}$ sample direction, different laminated displacement types are evident corresponding to $\hkl{100}$ type displacements, $\hkl{110}$ type displacements and higher order displacements along the tilt direction. The fine mixture of disconnection displacement types is mediated by chains of non-optimal displacements along the tilt axis which may be interpreted as microsliding vectors that maintain compatibility of the overall transformation. This migration pattern is consistent with the picture of roughening in \cite{han2018grain} as multiple disconnection islands nucleating in the plane of the GB. However, roughening of the displacement pattern (mixing of disconnection shuffles) is found to occur at a very fine scale with geometrically necessary non-optimal displacements. Such non-optimal displacements may impact the rate limiting step of roughened GB migration. The overall migration mechanism is not consistent with a zig-zag lamination pattern as observed for a different GB geometry in \cite{thomas2017}. This shows that disconnection displacement patterns may mix in multiple ways, depending on GB geometry and simulation conditions. The forward model enables analysis of multimodal migration data from atomistic simulations at the level of atomic displacements.

\section{Outlook and conclusion}\label{sec:conclusion}

In this work, we have used the principle of stationary action, re-written as an optimal transport problem, to predict permutations corresponding to grain boundary migration shuffles.
It was shown that this is equivalent to the min-shuffle hypothesis, and together with entropic regularization, was used to construct a computationally efficient potential-free forward model for enumerating GB migration mechanisms. 
The model has two primary advantages as a tool for interpreting and analyzing atomistic data.
First, the generality of the model allows it to be applied quickly to arbitrary CSL GBs and disconnection modes, which enables analysis of multi-modal data from atomistic simulations. 
Second, computations in the forward model are fast for systems with several hundred or fewer atoms and many calculations can easily be performed on a laptop. 
Because the model relies on well-established linear algebra routines, the model is readily adaptable to GPU acceleration.
Though out of the scope of the present work, this will facilitate extension of the model to large systems or parameter studies.

The model offers a new scientific perspective on the problem of GB migration by treating it as an optimal transport problem.
The connection to stationary action allows us to place quantitative bounds (via $\mathcal{D}$) on the range of validity of the the min-shuffle hypothesis transformation is expected to hold. 
The use of entropic regularization facilitates a deeper understanding of the role of non-optimal displacements in high-temperature, high driving force or high $\Sigma$ migration.
For energy jump driven migration at low and intermediate temperatures in the MD data, the min-shuffle transformation was found to dominate reorientation geometry. 
This suggests that, even with a non-negligible potential term, relative probability of optimal and non-optimal displacements depends primarily on GB geometry, with highly mobile GBs exhibiting a high fraction of optimal displacements under all simulation conditions. 

Importantly, the model offers an efficient and general way to quantify the role of translational microscopic degrees of freedom in mobility.
It was shown that the ``mobility-optimal'' shift vector is, in general, different from the orientation-dependent ``energy-optimal'' shift vector that minimizes the grain boundary energy. This may explain the connection (or lack thereof) between energy and mobility. The framework sheds new light on the manifestation of shear during grain boundary migration. Small sliding events separating metastable states may greatly impact the optimal transformation or non-optimal displacements associated with GB diffusion.



As with any reduced-order theory, the proposed forward model has a number of limitations and opportunities for improvement.
First, we strongly emphasize that the derivation of the model from stationary action should not be taken as proof that the non-equilibrium process of GB migration is fundamentally variational. 
The derivation of such a simple model required several strong simplifications: neglecting the potential term and driving force, approximating trajectories as linear (multi-stage or multi-hop events are neglected), and neglecting GB structure apart from a microscopic translation. We also caution against an overly physical interpretation of the entropic regularization parameter $\varepsilon$.
While it is tempting to identify this parameter as a thermodynamic temperature, it is nevertheless a heuristic observation that is not supported by a rigorous thermodynamic argument.
We also note that, while entropic regularization in the forward model is convenient for geometric segmentation of MD data, it does not consistently match the statistics of the MD data. 

In the future, the analysis in this work can be combined with detailed statistical analysis of atomic displacements during GB migration  \cite{zhang2006simulation,yan2010atomistic,mishin2015atomistic} to guide intuition for increasingly physics-based models of grain boundary motion that account for both sliding and GB diffusion. It would be very interesting to experimentally test the min-shuffle hypothesis via TEM experiments of low $\Sigma$ GB motion similar to \cite{wei2021direct}. 

\section*{Acknowledgments}

Work at CMU was supported in part by National Science Foundation grants GFRP-1252522 and DMR-1710186.
BR acknowledges computing support from the Extreme Science and Engineering Discovery Environment (XSEDE) allocation TG-PHY130007, which is supported by the National Science Foundation grant number ACI-1548562.

\bibliography{library}

\newpage

\appendix
\setcounter{figure}{0}

\section{NEB and geodesic costs for $\Sigma5\hkl{100}$}\label{sec:supplementary_information}

\begin{figure}[th]
    \centering\leavevmode
    \includegraphics[width=1\textwidth]{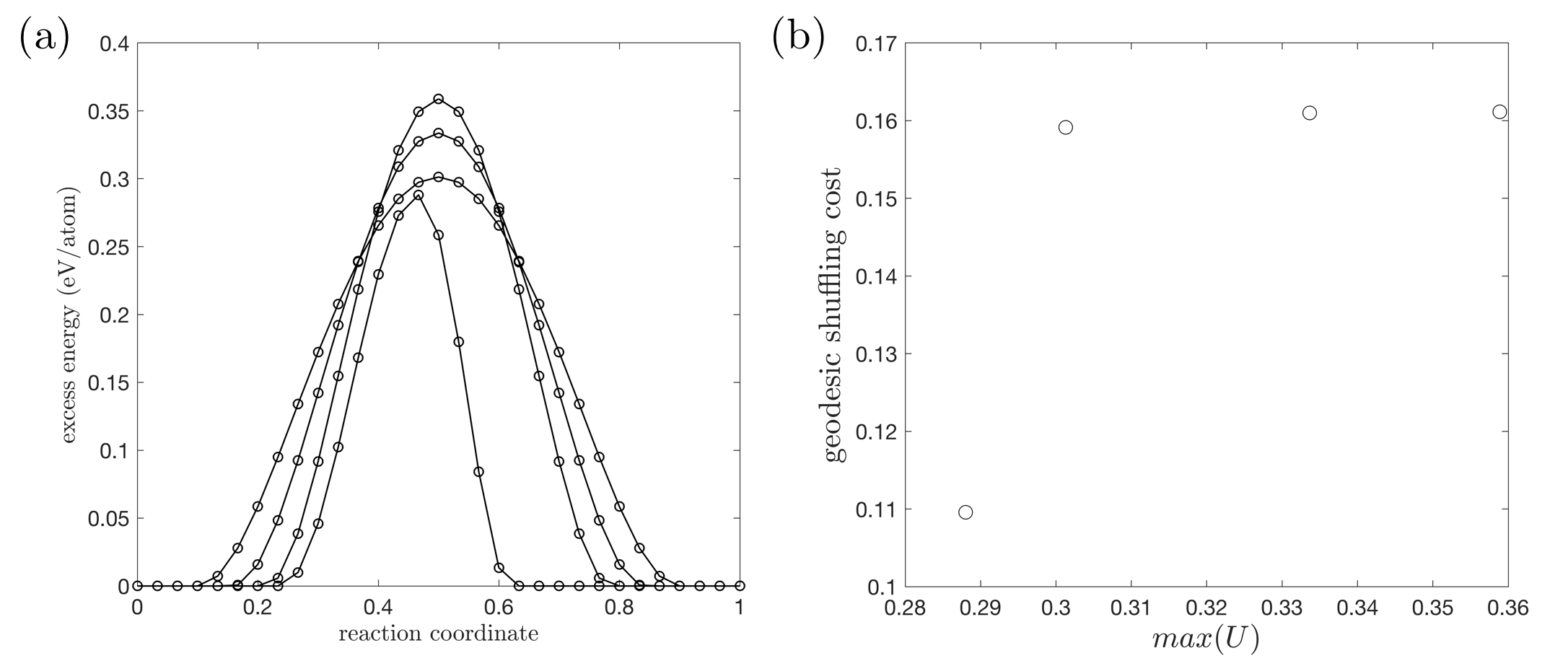}
    \caption{(a) NEB energy barriers and (b) geodesic costs for $k = 4$ lowest cost shuffle mappings for $\Sigma 5 \hkl{100}$ GB. Geodesic shuffling cost and barrier height are inversely correlated in this example.}
    \label{fig:S1}
\end{figure}

\end{document}